\documentclass{kapproc} % Computer Modern font calls
\let\footnote\savefootnote
\let\footnotetext\savefootnotetext 
\let\footnoterule\savefootnoterule 
\kluwerbib

\RequirePackage{graphicx}%
\RequirePackage{epsf}%
\input{psfig}

\def\lya{\ifmmode {\rm Ly}\alpha~ \else Ly$\alpha$~\fi}
\def\lyb{\ifmmode {\rm Ly}\beta~ \else Ly$\beta$~\fi}
\def\lyg{\ifmmode {\rm Ly}\gamma~ \else Ly$\gamma$~\fi}
\def\civ{\ifmmode {\rm C}\,{\sc iv}~ \else C\,{\sc iv}~\fi}
\def\civn{\ifmmode {\rm C}\,{\sc iv}~ \else C\,{\sc iv}\fi}
\def\cvn{\ifmmode {\rm C}\,{\sc v}~ \else C\,{\sc v}\fi}
\def\cvin{\ifmmode {\rm C}\,{\sc vi}~ \else C\,{\sc vi}\fi}
\def\nvn{\ifmmode {\rm N}\,{\sc v}~ \else N\,{\sc v}\fi}
\def\nvin{\ifmmode {\rm N}\,{\sc vi}~ \else N\,{\sc vi}\fi}
\def\nviin{\ifmmode {\rm N}\,{\sc vii}~ \else N\,{\sc vii}\fi}

\def\oi{{{\rm O}\,\hbox{{\sc i}}~}}

\def\ovi{{{\rm O}\,\hbox{{\sc vi}}~}}
\def\ovii{{{\rm O}\,\hbox{{\sc vii}}~}}
\def\oviii{{{\rm O}\,\hbox{{\sc viii}}~}}

\def\ovin{{{\rm O}\,\hbox{{\sc vi}}}}
\def\oviin{{{\rm O}\,\hbox{{\sc vii}}}}
\def\oviiin{{{\rm O}\,\hbox{{\sc viii}}}}
\def\neix{\ifmmode {\rm Ne}\,{\sc ix}~ \else Ne\,{\sc ix}~\fi}
\def\nex{\ifmmode {\rm Ne}\,{\sc x}~ \else Ne\,{\sc x}~\fi}
\def\hi{\ifmmode {\rm H}\,{\sc i}~ \else H\,{\sc i}~\fi}
\def\kms{\rm\,km\,s^{-1}}
\def\hubunits{\rm\,km\,s^{-1}\,Mpc^{-1}}
\def\K{\,{\rm K}}
\def\cm{{\rm cm}}
\def\chandra {{\it Chandra}~}

\def\etal   {{\it et~al.}~}

\begin{document}

\articletitle{X-ray Observations of the Warm-Hot Intergalactic Medium}

\author{Smita Mathur, David H. Weinberg, \& Xuelei Chen}

\affil{The Ohio State University}
\email{smita@astronomy.ohio-state.edu}

%% optional, to supply a shorter version of the title for the running head:
\chaptitlerunninghead{X-ray Observations of WHIM}

\begin{abstract}
We present \chandra observations that provide the most direct
evidence to date for the pervasive, moderate density, shock-heated
intergalactic medium predicted by leading cosmological scenarios. We
also comment briefly on future observations with Constellation-X.
\end{abstract}

% begin document here
\section{Introduction}
Much of our knowledge of the intergalactic medium (IGM) comes from the
rest-frame UV line absorption that it imprints on the spectra of
background quasars: the \lya forest of neutral hydrogen and associated
metal lines such as \civ and \ovi.  The \lya forest thins out at low
redshift, and cosmological simulations predict that the continuing
process of structure formation heats a substantial fraction of
intergalactic gas to temperatures where it produces little hydrogen
\lya absorption, and where the dominant ionization stages of heavier
elements have absorption transitions at X-ray wavelengths rather than
UV.  One of the few prospects for detecting this low density,
shock-heated gas is via the ``X-ray forest'' of absorption lines it
should produce in quasar spectra (Hellsten, Gnedin, \&
Miralda-Escud\'{e}(1998)). Here we discuss X-ray forest searches, with
focus on absorption towards H1821+643 ($z=0.297$) using a high
resolution ($\lambda/\Delta\lambda \approx 500$) spectrum obtained
with the \chandra X-ray Observatory.

The baryon density implied by big bang nucleosynthesis and the
estimated primordial deuterium abundance, $\Omega_{\rm BBN} \approx
0.04 h_{70}^{-2}$ (Burles \& Tytler (1998); here $h_{70} \equiv
H_0/70\;\hubunits$), exceeds the density of baryons in known stars and
X-ray emitting gas by roughly an order of magnitude (Fukugita, Hogan,
\& Peebles 1998).  The lower density regime of the ``warm-hot
intergalactic medium'' (WHIM, Cen \& Ostriker, 1999) could constitute
a major fraction of the ``missing'' low redshift baryons.
Hydrodynamic cosmological simulations predict that 30-50\% of the
baryons reside in this phase at $z=0$ (Dav{\' e} et al. 2001).  HST
and FUSE detections of \ovi ($\lambda\lambda 1032, 1038$\AA)
absorption lines towards H1821+643 (Tripp, Savage, \& Jenkins 2000,
hereafter TSJ) offer a tantalizing hint of this baryon reservoir.
Adopting conservative assumptions of [O/H]$=-1$ and an \ovi ionization
fraction $f$(\ovin)=0.2 (which is close to the maximum in photo- or
collisional ionization), TSJ conclude that the gas associated with
these weak \ovi absorbers accounts for $\Omega_b \approx 0.004
h_{70}^{-1}$ of the cosmic baryon budget, comparable in total mass to
all other known low redshift components combined.  For most reasonable
assumptions about the physical conditions of this absorbing gas, the
dominant ionization state should be \ovii or \oviiin, and $f(\ovin)$
lower than 0.2, implying a substantially higher baryon fraction.

\section{\chandra observations of H1821+643}

We chose H1821+643 for our X-ray forest search for two reasons: (1) it
is one of the brightest X-ray quasars at moderate redshift, and (2) it
has been studied carefully for intervening \ovi absorption (TSJ;
(Oegerle \etal 2000). The latter is important because even long
observations with \chandra gratings were expected yield a low S/N
spectrum, and searching at redshifts with known \ovi absorption allows
one to adopt a lower effective threshold for significant detection,
greatly increasing the odds of success. Furthermore, detections of or
upper limits on X-ray absorption provide new constraints on the
physical conditions of the six known \ovi absorbers (z1=0.26659,
z2=0.24531, z3=0.22637, z4=0.22497, z5=0.21326 , z6=0.12137).

%We observed H1821+643 with \chandra LETG for 480 ksec; the 21--28 \AA~
%spectrum is shown in figure 1. 
%With this observation, we expected to
%detect the strongest of the TSJ systems, and we would detect the next
%two strongest systems at the $2-3\sigma$ level. What we found,
%however, was a surprise; we find only a $\sim 1\sigma$ signal from the
%strongest TSJ system, but we obtain $\approx 2\sigma$ detections of
%\ovii from the next strongest system and of \ovii and \oviii from the
%\ovi system discovered by \cite{oegerle00}. This immediately implies
%strong variations from system to system in $f(\ovin)$. We also find
%features that could correspond to oxygen absorption at other redshifts
%close to the detected systems (within 1000$\kms$), and a possible
%signature of \neix absorption from one system.

\begin{figure} [h]
\psfig{file=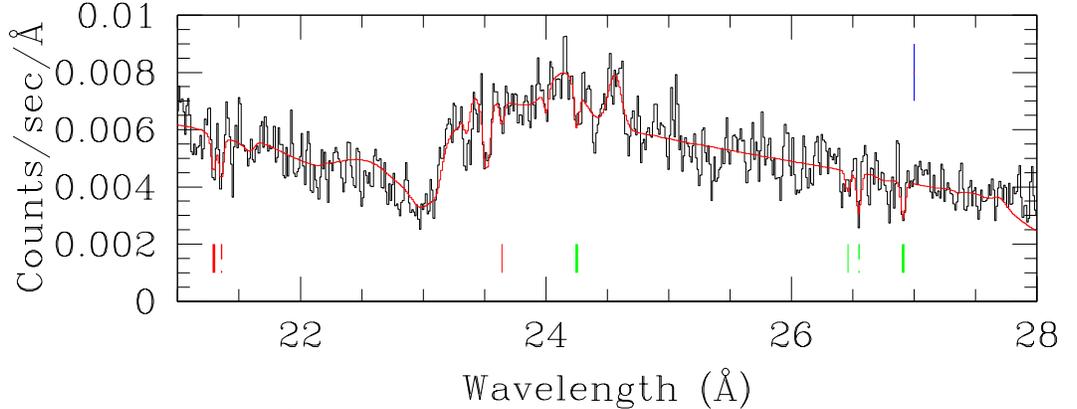,width=6in}
\vspace*{-3.5in}
\caption{The $21-28$\AA\ region of the observed spectrum. The red line
running through the spectrum delineates the continuum $+$ absorption
line model fit to the data.  The \ovii lines are marked in green and
\oviii lines in red. The solid tickmarks indicate lines at the known
\ovi redshifts, with thick lines indicating ``detected'' systems,
whose significance is $\ga 2\sigma$.  The dashed tickmarks indicate
the candidate ``new'' systems, which are features of comparable
strength within $1000\;\kms$ of known \ovi redshifts.  The broad \oi
absorption line at $\sim 23.5$\AA\ is from the Galaxy and the O-K edge
at $\sim 23$\AA\ is from the instrument as well as the Galaxy. The
blue bar in the upper right corner of the figure represents a typical
error-bar.}
\end{figure}

We observed H1821+643 with \chandra LETG for 480 ksec; the 21--28 \AA~
spectrum is shown in figure 1.
The \chandra spectrum was fitted with a smooth continuum plus a
Gaussian absorption line at the $\ovin \lambda$21.602\AA\ wavelength
for each of the \ovi redshifts z1 to z6 (see Mathur \etal 2002 for
details of observations and analysis). This fit yielded $2-3\sigma$
detections at z2 and z6. Applying the same procedure at the expected
wavelengths of the \oviii $\lambda$18.969\AA\ line yielded a
significant detection only at z6.

The existence of features with $\ga 2\sigma$ significance at several
wavelengths predicted {\it a priori} on the basis of \ovi absorption
suggests that we have indeed detected X-ray forest lines from highly
ionized oxygen in these systems.  However, we cannot rule out the
possibility that the coincidence between observed features and known
absorption redshifts is, in fact, just a coincidence. We calculate the
probability of a coincidence, in the absence of a true physical
correlation, to be about 5\%.  Since this probability is not extremely
small, we will be cautious in interpreting our observations,
considering both the possibility that we have true detections of
several X-ray forest absorbers and the possibility that we have only
upper limits. 

\section{Implications}

We convert the observed line equivalent widths to corresponding column
densities for each ion species assuming optically thin gas.  The z2
system has an inferred \ovii column density of $3.9 \pm 1.7 \times
10^{15}\cm^{-2}$, while the z6 system has $N(\oviin)=2.8 \pm 1.5\times
10^{15}\cm^{-2}$ and $N(\oviiin)=6.7 \pm 3.5 \times
10^{15}\cm^{-2}$. Upper limits for undetected systems are typically
$\sim 1.9\times 10^{15}\cm^{-2}$ for \ovii and $\sim 3\times
10^{15}\cm^{-2}$ for \oviii. Using these, we estimate
$f(\oviin)/f(\ovin)$ and $f(\oviiin)/f(\ovin)$ for the six observed
systems and compare them to theoretical predictions (Chan \etal
2002). Figure 2 shows tracks in $f(\oviiin)/f(\ovin)$ vs.\
$f(\oviin)/f(\ovin)$ plane for the intergalactic medium for a range of
temperatures and densities. The dot-dashed line is for pure collisional
ionization while the other contours include photoionization by the UV
and the soft X-ray background, and correspond to overdensities
$\delta_b (1+z)^3 =1$, 10, $10^2$, and $10^3$. The right panel shows
lines of constant temperature. Comparing the two panels shows that
$f(\oviin)/f(\ovin)$ is primarily a diagnostic of gas temperature,
while $f(\oviiin)/f(\ovin)$ constrains the gas density for a given
value of $f(\oviin)/f(\ovin)$.  This behavior reflects the competing
roles of photoionization and collisional ionization, with the latter
being more important for higher temperatures, higher densities, and
lower ionization states.  The filled points with $1\sigma$ error
crosses mark the two detected \ovi systems at z2 and z6. For the
remaining four systems, we plot arrows showing the $1\sigma$ upper
limits on $f(\oviin)/f(\ovin)$ and $f(\oviiin)/f(\ovin)$. 

\begin{figure} [h]
\centerline{
\epsfxsize=5.0truein
\epsfbox[65 450 550 720]{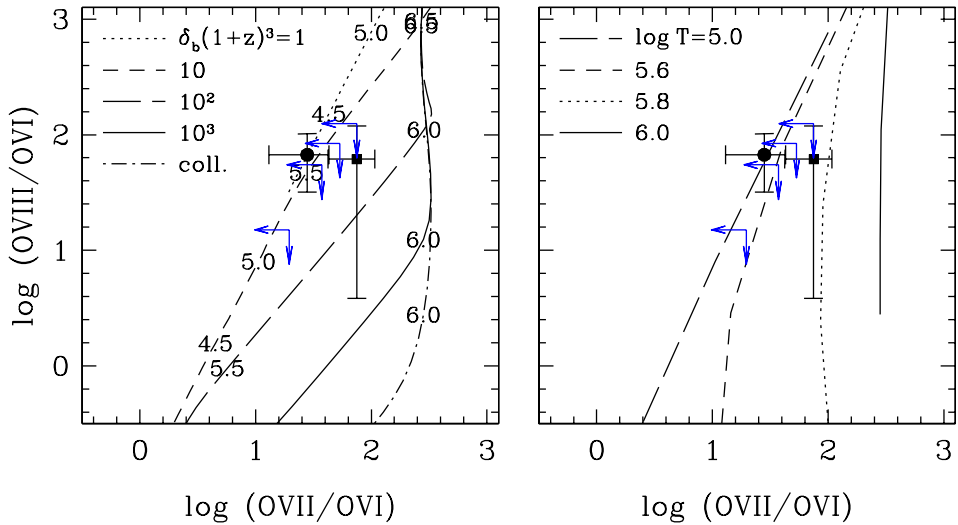}
}
\caption{Constraints on the physical state of the known \ovi
absorbers. Left: Curves show the tracks in the $f(\oviiin)/f(\ovin)$
vs.\ $f(\oviin)/f(\ovin)$ plane, based on photo- and
collisional-ionization calculations.  Numbers along these curves
indicate $\log T$ in degrees Kelvin. Points with $1\sigma$ error bars
show the detected systems at z2 (square) and z6 (circle).  Blue arrows
indicate $1\sigma$ upper limits for the remaining four systems, at z4,
z1, z5, and z3 (bottom to top).  Right: Same, but with tracks of
constant temperature.  }
\end{figure}

At the $\sim 1\sigma$ level, our measurements and limits have a number
of interesting implications.  First, there are variations in the ion
ratios from system to system. Second, the upper limits on the
undetected systems restrict their locations in the temperature-density
plane. For example, the z4 system must have $T \la 10^{5.5}\K$, and if
it is close to this temperature, it must have $\delta_b \ga 5$.
Third, the overdensities implied by the best-fit line parameters of
the detected systems at z2 and z6 are significantly below the values
$\delta_b \sim 100$ corresponding to virialized systems.  In physical
terms, the co-existence of detectable amounts of \ovin, \oviin, and
\oviii requires that photoionization play a central role in
determining the abundance ratios, which is possible only if the
density is fairly low.

\section{The baryon budget}

The major uncertainty in determining the contribution of \ovi systems
to the cosmic baryon budget is $f(\ovin)$, and this quantity is what
we can estimate using the X-ray observations. We treat all of our
$2\sigma$ detections as real, but where we have only $1\sigma$
measurements or upper limits, we conservatively assume the lowest
\ovii and \oviii column densities consistent with the physical
expectation that $f(\ovin)\leq 0.2$. This calculation yields $\langle
[f(\ovin)]^{-1} \rangle = 32 \pm 9$, substantially higher than the
conservative value of 5.0 that TSJ assumed.  Combined with the TSJ
result, this ratio implies $\Omega_b(\ovin) = 0.028 \pm 0.008 \;
h_{70}^{-1}$ for $[{\rm O/H}]=-1$, which substantially exceeds the
contribution of any other known low redshift baryon component,
representing an appreciable fraction of the baryons predicted by BBN.

\section{A note of caution}

The implications discussed above are only as compelling as the
detections themselves. Constraints on the physical conditions of the
IGM at the $2\sigma$ level are rather weak.  The conclusion that $T <
10^6\K$, with a tighter upper limit for stronger absorbers, holds
robustly, but stronger physical constraints on the \ovi systems at the
$2\sigma$ level require higher S/N than our present observations
afford. However, we can still derive an upper limit on $\langle
[f(\ovin)]^{-1} \rangle$, even if the $2\sigma$ lines are not real
detections. The result is $\langle [f(\ovin)]^{-1} \rangle < 60$ at
$1\sigma$ and $\langle [f(\ovin)]^{-1} \rangle < 79$ at $2\sigma$.  In
combination with the TSJ  numbers, even the
$1\sigma$ upper limit is consistent with $\Omega_b(\ovin) \approx
\Omega_{\rm BBN}$.  If we had obtained null results for all of the
\ovi systems, then the upper limit on $\Omega_b(\ovin)$ would have
come out well below $\Omega_{\rm BBN}$.

\section{Discussion and future observations} 

Our results show that definitive measurements of X-ray forest
absorption are extraordinarily difficult. Fang \etal (these
proceedings) claim a 4.5$\sigma$ detection of an intervening \oviii
system towards PKS 2155-304. However, XMM-Newton observations
(Rasmussen \etal, these proceedings) rule out any detection at the
reported significance. Thus, even with the great technological advance
of \chandra and XMM-Newton, we do not have  fully convincing evidence
yet of any X-ray forest absorption beyond the local group. Mapping
the warm-hot IGM appears to be {\it Constellation-X} science. A 500
ksec observation with {\it Constellation-X} grating will result in an
equivalent width limit of 2.8 m\AA\ ($5\sigma$). If the grating
resolution is further improved to R=3000, then X-ray lines as weak as
1.4m\AA\ will be detected. The higher resolution is important because
there may not be tall trees in the X-ray forest.

The tantalizing but still ambiguous hints of X-ray forest absorption
in current observations are frustrating to live with for half a decade
or more before the launch of {\it Constellation-X}. In the mean time
there are a couple of options for progress. We can wait for a blazar to
flare up; a good S/N spectrum may then be possible  in a
reasonable time. However, the number of X-ray forest systems with
column densities large enough to be observable with \chandra gratings
is expected to be only about one per unit redshift. This makes success
 unlikely. The best way forward may be a deep (1Ms) observation
of H1821+643. H1821+643 remains the best target for
such an investigation because of its X-ray brightness and well studied
\ovi absorption, and because the {\it Chandra} spectrum presented here
provides 500 ksec of existing data and clear objectives for a future
observation.

% Deluxetable will work with this style file
% Plotone, plottwo, and plotfiddle should all work for plotting

%enter bibliography here

\end{document}